\documentclass{aa}

\usepackage{txfonts}
\usepackage{graphicx}

\begin{document}

\title{Analytical computation of the off-axis effective area of grazing incidence X-ray mirrors}

\author{D. Spiga, V. Cotroneo, S. Basso, P. Conconi}

\institute{INAF/Osservatorio Astronomico di Brera, Via E. Bianchi 46, I-23807, Merate (LC) - Italy\\ \email{daniele.spiga@brera.inaf.it}}

\date{Received 9 June 2009 / Accepted 20 June 2009 }

\abstract 
{}
{Focusing mirrors for X-ray telescopes in grazing incidence, introduced in the 70s, are characterized in terms of their performance by their imaging quality and effective area, which in turn determines their sensitivity. Even though the on-axis effective area is assumed in general to characterize the collecting power of an X-ray optic, the telescope capability of imaging extended X-ray sources is also determined by the variation in its effective area with the off-axis angle. The effective area, in general, decreases as the X-ray source moves off-axis, causing a loss of sensitivity in the peripheral regions of the telescope's field of view.}
{The complex task of designing optics for future X-ray telescopes entails detailed computations of both imaging quality and effective area on- and off-axis. Because of their apparent complexity, both aspects have been, so far, treated by using ray-tracing routines aimed at simulating the interaction of X-ray photons with the reflecting surfaces of a given focusing system. Although this approach has been widely exploited and proven to be effective, it would also be attractive to regard the same problem from an analytical viewpoint, to assess an optical design of an X-ray optical module with a simpler calculation than a ray-tracing routine. This would also improve the efficiency of optimization tasks when designing the X-ray optical modules. In this paper, we thereby focused on developing analytical solutions to compute the off-axis effective area of double-reflection X-ray mirrors.}
{We have developed useful analytical formulae for the off-axis effective area of a double-reflection mirror in the double cone approximation, requiring only an integration and the standard routines to calculate the X-ray coating reflectivity for a given incidence angle. The computation is easily applicable also to Wolter-I mirrors (such as those of NeXT, NuSTAR, HEXIT-SAT, IXO) and the approximation improves as the f-number of the mirror increases. Algebraic expressions are provided for the mirror geometric area, as a function of the off-axis angle. Finally, the results of the analytical computations presented here are validated by comparison with the corresponding predictions of a ray-tracing code.} 
{}

\keywords{Telescopes -- Methods: analytical}
\titlerunning{Analytical computation of the off-axis Effective Area}
\authorrunning{D. Spiga et al.}
\maketitle

\section{Introduction}
\label{Intro}
X-ray telescopes have been equipped with focusing optics since the 70s to endow them with imaging capabilities and concentration properties that enhance their sensitivity. The effective area of the optics is one of the most important parameter determining the minimum detectable flux: to date, the X-ray telescope with the largest effective area in the soft X-ray band ($<$~10~keV) is Newton-XMM with 1450~cm$^2$ per module at 1 keV, on-axis (Gondoin~et~al.~\cite{Gondoin2}). Beyond 10 keV, the effective area and the sensitivity of all focusing X-ray telescopes drop off suddenly, because of the very low reflectivity of single-layer coated mirrors at the grazing incidence angles in use. The situation is expected to change with the launch of imaging hard ($>$ 10 keV) X-ray telescopes such as NuSTAR (Koglin~et~al.~\cite{NuSTAR}), NeXT (Takahashi~et~al.~\cite{Takahashi}; Ogasaka~et~al.~\cite{Ogasaka}), HEXIT-SAT (Pareschi~\cite{Pareschi04}), and IXO (formerly XEUS, Parmar~et~al.~\cite{Parmar}; Kunieda~et~al.~\cite{Kunieda}) that will extend the performances of presently-operating X-ray telescopes beyond 10 keV, by means of shallow incidence angles ($<$~0.25~deg) and wideband multilayer coatings (Joensen~et~al.~\cite{Joensen}; Tawara~et~al.~\cite{Tawara}).

On the other side, the study of extended X-ray sources, such as galaxy clusters, and even cosmological surveys would benefit greatly from an increase in the {\it field of view} of X-ray telescopes. For example, eROSITA (Predehl~et~al.~\cite{Predehl}) will provide a wide field of view of 61 arcmin in diameter (Friedrich~et~al.~\cite{Friedrich}) by using double reflection mirrors. Previous mission concepts, such as WFXT (Citterio~et~al.~\cite{Citterio}) and WFI onboard EDGE (Piro~et~al.~\cite{Piro}), have been proposed to increase the imaging quality of off-axis sources, by adopting polynomial profiles (Burrows~et~al.~\cite{Burrows}; Conconi \& Campana~\cite{CampanaConconi}). This type of mirror design provides a higher resolution for off-axis X-ray sources than the Wolter's, at the expense of a small degradation of the HEW on-axis. 

Nevertheless, not only the angular resolution, but also the effective area of grazing-incidence X-ray mirrors is known to be degraded as a source moves off-axis, due to geometrical vignetting and the variation in the incidence angles on mirrors. This can have important consequences for the observation, since the sensitivity of the telescope over its field of view can be severely compromised. Moreover, for telescopes with modules flying in formation (as was to have been the case for SIMBOL-X, Pareschi~et~al.~\cite{Pareschi08}), oscillations at random of the optical module might cause an unpredictable decrease in the effective area. Possible solutions to this problem have been studied (Cotroneo~et~al.~\cite{Cotroneo2009}), although this example highlights the importance of the theoretical prediction of the effective area, on- and off-axis, already at the design stage of the optical module development for an X-ray telescope. 

Another situation in which the theoretical computation of the effective area is necessary occurs whenever X-ray mirrors are calibrated using on-ground facilities such as PANTER (Brauninger~et~al.~\cite{Brauninger}; Freyberg~et~al.~\cite{Freyberg}), where the X-ray source is located at a large, but definitely non-astronomical, distance from the mirror. In such conditions, the measured effective area is affected by the imperfect collimation of the incident beam: the correct data interpretation must account for this effect (see e.g., Gondoin~et~al.~\cite{Gondoin2}), on- and off-axis, to reliably reconstruct the mirror's effective area for an astronomical X-ray source over all the field of view of the telescope.

While the prediction of the on-axis effective area for a double-reflection, grazing-incidence mirror is rather simple, it becomes difficult in general for a source off-axis. Such evaluations have been completed so far using ray-tracing codes (see e.g., Mangus~\&~Underwood~\cite{Mangus}; Zhao~et~al.~\cite{Zhao}), starting from the optical design of mirrors and the structure of the reflective coating. These codes are in general very efficient, but time-consuming, computationally intensive, and affected by statistical errors related to the number of rays that can be traced. It would therefore be beneficial to develop a method for computing the effective area {\it analytically}, which would ease the assessment of the effective area and, among other things, the optical design of a wide field X-ray telescope. 

Few suitable analytical tools have been developed to date. Van~Speybroeck and Chase~(\cite{VanSpeybroeck}) found -- by means of ray-tracing -- that the {\it geometric}, collecting area of a Wolter-I mirror decreases with the off-axis angle $\theta$ of the source, with respect to the on-axis geometric area $A_{\infty}(0)$, according to
\begin{equation}
	A_{\infty}(\theta) = A_{\infty}(0)\,\left(1-\frac{2\theta}{3\alpha_0}\right),
\label{eq:SC_formula}
\end{equation}
where $\alpha_0$ is the incidence angle for a source on-axis. However, we lack a {\it general, analytical method to derive the effective area of an X-ray mirror} with a given reflective coating, as a function of the off-axis angle of the X-ray source. 

In this paper, we present a solution to that problem. We develop an analytical approach that can be applied to double cone grazing-incidence X-ray mirrors and, with reasonable accuracy, to Wolter-I mirrors (unless the f-number is small). The limits of this approximation are discussed in Sect.~\ref{DC_WI}. In Sect.~\ref{EffArea}, we derive general integral formulae (such as Eq.~(\ref{eq:Aeff_fin_offaxis})) to compute the off-axis effective area for a double-reflection X-ray mirror with shallow incident angles, for {\it any} reflective coating. As a particular case, in Sect.~\ref{Geometric} we obtain some algebraic expressions for the geometric area and verify that the well-known Eq.~(\ref{eq:SC_formula}) can be derived as a particular case. In Sect.~\ref{Comp}, the predictions of the analytical approach are validated, for some particular cases, by means of a comparison with the outputs of a ray-tracing routine. The results are briefly discussed in Sect.~\ref{Final}.

We note that we assume that the off-axis mutual obstruction of mirrors in densely nested mirror assemblies has a negligible effect. Therefore, the results are valid for either isolated double cone or Wolter-I mirrors, or for mirror modules known to be negligibly obstructed, such that their effective area simply equals the sum of the contributions of the individual mirrors. The quantification of the off-axis obstruction in mirror assemblies will be considered in future.

\section{The double cone approximation in the computation of the effective area of a Wolter-I mirror}
\label{DC_WI}
\begin{figure}
	\resizebox{\hsize}{!}{\includegraphics{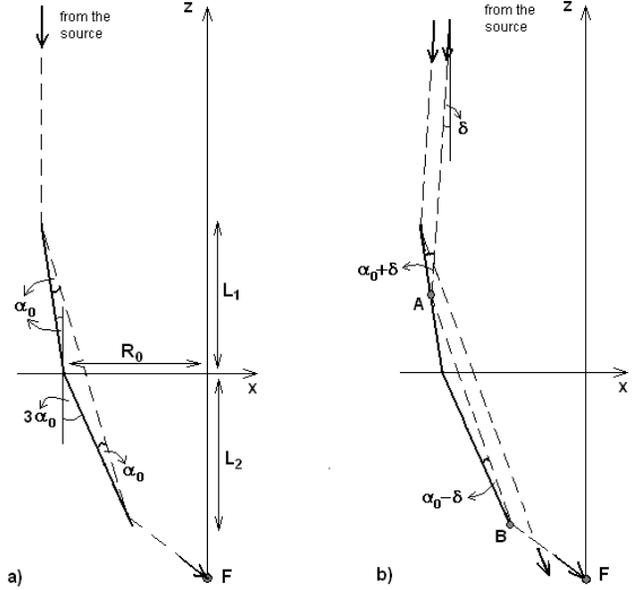}}
	\caption{Meridional section of a Wolter-I mirror, here in the case $L_2 \geq L_1$: a) source on-axis, at astronomical distance; b) source still on-axis, but located at finite distance (all angles are greatly exaggerated). }
\label{fig:mirror_section}
\end{figure}

We consider, in a preliminary way, a grazing-incidence Wolter-I mirror, and an on-axis photon source (Fig.~\ref{fig:mirror_section}). The optical axis is aligned with the $z$ axis. We define $R_{\mathrm M}$ to be the radius at the parabolic end (i.e., the maximum radius), $R_0$ the radius at $z=0$ (the {\it intersection plane}), $R_m$ the radius at the hyperbolic end (i.e., the minimum radius), $F$ the focal point, and $f$ the distance of $F$ from the intersection plane (i.e., the focal distance). In general, we refer to ``primary'' and ``secondary" segments, instead of ``parabola" and ``hyperbola". We denote with $L_1$ the primary segment length along the $z$ axis, and $L_2$ that of the secondary. The polar coordinate is $\varphi$.

Because of the surface curvature, the incidence angles on the two surfaces vary in general with the $z$ coordinate. We define $\alpha(z)$ to be this angle on the primary ($0<z<L_1$) mirror segment, and on the secondary ($-L_2<z<0$) mirror for a source on the optical axis, at infinite distance. In general (Van~Speybroeck~and~Chase~\cite{VanSpeybroeck}), the incidence angle of rays close to $z$ = 0, $\alpha_0$, is the same for both surfaces. Therefore, we have the well-known relation 
\begin{equation}
	R_0 = f \tan(4\alpha_0).
\label{eq:Rf}
\end{equation}
Since $\alpha_0$ is shallow, we hereafter assume that the $\tan(4\alpha_0)$ function can be approximated by $4\alpha_0$ itself. 

To obtain an analytical expression for the effective area of a Wolter-I mirror when a source is off-axis by an angle $\theta$, we have to find algebraic expressions for:
\begin{enumerate}
	\item{the variation in the incidence angles, on which the mirror reflectivities depend, over the primary and secondary mirror segment surfaces; }
	\item{the collecting area of the primary segment;}	
	\item{the fraction of rays reflected by the primary segment that are incident on the secondary.}
\end{enumerate}
To compute the effective area for a source off-axis, we begin with the prototypical case of a mirror with a source on-axis, but at a finite, although very large, distance $D$, which is the usual configuration for on-ground calibration facilities. All rays impinge the primary segment within a meridional plane, and in the double cone approximation they are all incident at the same grazing angle, independently of $z$ and $\varphi$. The finite distance of the source causes the beam to have a divergence at the primary segment, which can be assumed to be constant as long as $D \gg L_1$, with an half-aperture angle of $\delta \simeq R_0/D$. In this simplified case, every point of the mirror sees the source {\it off-axis} by an angle $\delta$, regardless of $\varphi$. We demonstrate hereafter that the solution to these problems is simpler to express analytically if the profile of the mirror can be approximated with a double cone, if we are interested only in the effective area. In contrast, the curvature of mirrors along the axis is essential for their angular resolution, but we do not consider this aspect here.

In the remainder of this section, we quantify the errors caused by the substitution of a Wolter-I profile with a double cone, by keeping $R_0$, $\alpha_0$, $L_1$, and $L_2$ constant. The problems to be faced are related to the system geometry, rather than to the absolute size of the mirrors. Hence, it is convenient to report the results of this section in terms of both the f-number, $f\# = f/(2R_0)$, instead of $f$ itself, and $L_1' = L_1/(2R_0)$, $L_2' = L_2/(2R_0)$, which are normalized mirror lengths. Using Eq.~(\ref{eq:Rf}), $f\#$ can be written as 
\begin{equation}
	f\# \simeq \frac{1}{8\alpha_0}.
	\label{eq:f_num}
\end{equation}
 We mention that the true focal length of a double cone is slightly longer than that of a Wolter profile with the same $R_0$ and $\alpha_0$, because of the different focusing properties. In all cases, we always refer to $f$ as the focal length of the corresponding Wolter-I mirror, to ensure that Eq.~(\ref{eq:Rf}) retains its validity.

\subsection{The incidence angles}\label{inc_angles}
For a double cone profile, the incidence angles for an on-axis source at infinity are constant along a meridional plane, and equal to $\alpha_0$. For a Wolter-I profile, the slope of the mirror profile changes slowly with the $z$ coordinate. Because of the concavity of profiles, $\alpha(z) \leq \alpha_0$ on the primary segment, and $\alpha(z) \geq \alpha_0$ on the secondary segment. We assume for simplicity that $L_1=L_2=L$, and define the (positive) angle variations $\Delta\alpha_1 \simeq \alpha_0-\alpha(+L)$ and $\Delta\alpha_2 \simeq \alpha(-L)-\alpha_0$. For a Wolter-I profile, the ratio $\Delta\alpha_2/ \Delta\alpha_1$ varies between 1 and 2, going from large to short $f\#$. For this estimation, we assume that $\Delta\alpha_2 \approx \Delta\alpha_1$ and we simply denote their value with $\Delta\alpha$.

Now consider a ray from an on-axis source at infinity (Fig.~\ref{fig:mirror_section}a). After being reflected firstly at $z = +L$, it undergoes the second reflection at $z \simeq -L$, at a radial coordinate $R_m$, and is focused at $z = -f$. The total photon deflection at $z =-L$ is $2(\alpha_0+\Delta\alpha_1)+2(\alpha_0+\Delta\alpha_2)$; therefore,
\begin{equation}
	R_m = 4(f-L)(\alpha_0+\Delta\alpha).
	\label{eq:angle_var0}
\end{equation}
For small $\alpha_0$ we also have that
\begin{equation}
 	R_m \simeq R_0-3L\alpha_0.
 \end{equation}
By comparing the last two equations, using Eqs.~(\ref{eq:Rf}) and~(\ref{eq:f_num}), with the definition of $L_1' = L_2' = L'$, we derive the maximum variation in the incidence angles over the mirror profiles,
\begin{equation}
	\frac{\Delta\alpha}{\alpha_0} \approx \frac{L'}{4(f\#-L')}\label{eq:angle_var1}.
\end{equation}
This equation provides the error in the incidence angles introduced by the double cone approximation. Such an error is computed for some practical cases and reported in Table~\ref{tab:DC_toler}. For example, we note that for NeXT-HXT and SIMBOL-X the error is only 0.4\%. This means that, assuming $\alpha_0 \simeq 0.27$ deg at most, the angular variation along the profile is $\Delta \alpha \simeq $ 4 arcsec, which has almost no practical influence on the reflectivity of wideband multilayers. 

We now consider the source on-axis at finite distance, with $\delta <\alpha_0$. For a double cone (Fig.~\ref{fig:mirror_section}b), the incidence angle on the primary segment becomes $\alpha_1 = \alpha_0+\delta$, for all $\varphi$. Similarly, the incidence angle on the secondary segment is also constant and equal to $\alpha_2~=~\alpha_0-\delta$, provided that $\alpha_0>\delta$. Within the approximation of Eq.~(\ref{eq:angle_var1}), this is also true for a Wolter-I mirror.

\begin{figure}
	\resizebox{\hsize}{!}{\includegraphics{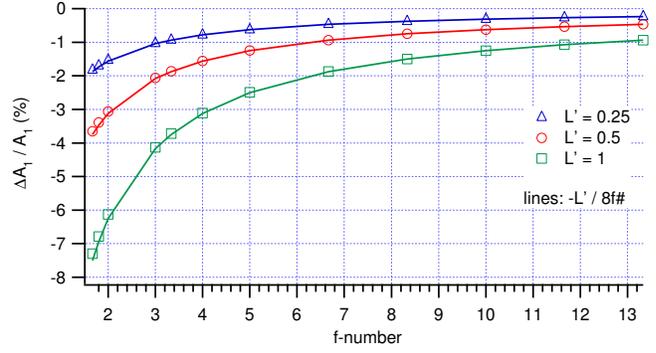}}
	\caption{Percent accuracy of the primary segment cross-section area, as seen by a source on-axis from infinity, in approximating a Wolter-I profile with a double cone. }
\label{fig:max_diam}
\end{figure}

\begin{table}
	\caption{Some examples of approximation introduced by the double cone geometry (optical parameters after Pareschi~et~al.~\cite{Pareschi08}, Ogasaka~et~al.~\cite{Ogasaka}, Koglin~et~al.~\cite{NuSTAR}, Gondoin~et~al.~\cite{Gondoin1}).}
	\label{tab:DC_toler}
	\centering
	\begin{tabular}{c c c c c}
	\hline \hline
	&SIMBOL-X & NeXT-HXT & NuSTAR & XMM\\
	\hline
	\\
	$f$& 20 m & 12 m & $\sim$10 m & 7.5 m\\
	\\
	$R_{0, {\mathrm max}}$& 324 mm & 225 mm & $\sim$169 mm & 346 mm\\
	\\
	$L$& 300 mm & 200 mm & 200 mm & 300 mm\\
	\hline
	\\
	$\frac{\displaystyle\Delta\alpha}{\displaystyle\alpha_0}$ & 0.4\% & 0.4 \% & 0.5\% & 1\%\\
	\\
	$\frac{\displaystyle\Delta A_1}{\displaystyle A_1}$& -0.18\% & -0.20\% & -0.25\% & -0.5\%\\
	\\
	$\frac{\displaystyle\Delta V}{\displaystyle V}(\delta =0)$& 1.5\% & 1.7\% & 2\% & 4\% \\
	\hline
	\end{tabular}
\end{table}

\subsection{The maximum diameter}\label{primary_area}
In terms of effective area, another concern of approximating a Wolter-I profile with a double cone of the same $R_0$, $\alpha_0$, $L_1$, and $L_2$, is the different cross-section of the primary segment. For a double cone profile, the maximum diameter is $R_M = R_0+ \alpha_0L_1$. For a Wolter-I mirror, because of the profile concavity, the maximum diameter is smaller by an amount that depends on $L'$ and $f\#$. Consequently, the geometric area of the primary segment is smaller for a Wolter profile than for a double cone with the same size and $\alpha_0$. However, the (negative) error caused by the substitution of a Wolter-I profile with a double cone, defined as 
\begin{equation}
	\frac{\Delta A_1}{A_1} =: \frac{A_1(\mbox{Wolter})-A_1(\mbox{double cone)}}{A_1(\mbox{Wolter})},
	\label{eq:defdeltaA}
\end{equation}
is, in general, of a few percent as far as $L' < 1$ (see Fig.~\ref{fig:max_diam}). In addition, its magnitude decreases rapidly for increasing $f\#$, since it is expressed well by the empirical formula
\begin{equation}
	\frac{\Delta A_1}{A_1}\approx - \frac{L'}{8f\#},
	\label{eq:max_diam}
\end{equation}
of the same order of magnitude as the expression in Eq.~(\ref{eq:angle_var1}). This approximation is also computed for some telescopes and reported in Table~\ref{tab:DC_toler}.

For an on-axis source at a finite distance, the area of the circular corona between $R_M$ and $R_0- L_1 \delta$ should be considered: the area is thereby increased by the same term $\sim 2\pi R_0L_1\delta$, for both Wolter and double cone. Hence, the approximation of the area is even better than the expression in Eq.~(\ref{eq:max_diam}). 

\begin{figure}
	\resizebox{\hsize}{!}{\includegraphics{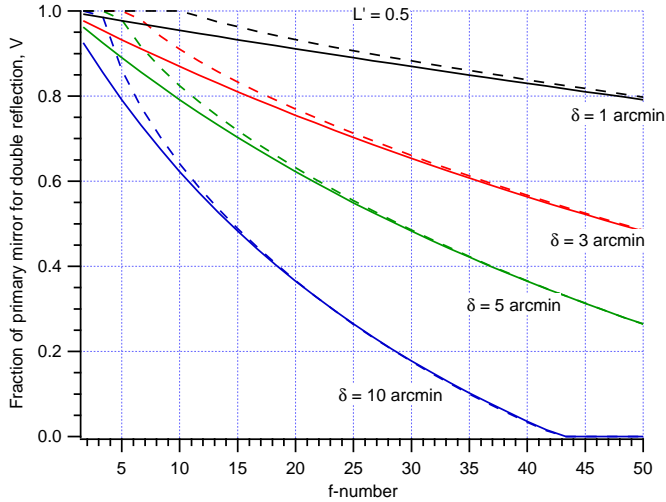}}
	\caption{Vignetting due to the finite distance of the source {\it on-axis}, as a function of the f-number, for $L'_1 = L'_2$ and 4 different values of $\delta$. Solid lines are computed assuming a Wolter-I profile, dashed lines a double cone (Eq.~(\ref{eq:vign_fnum})). The results for the double cone are completely independent of the choice of $L'$: those for the Wolter-I are computed assuming $L'~=~0.5$.}
\label{fig:fin_dist}
\end{figure}

\subsection{Geometric vignetting for double reflection}\label{vignetting}
We now consider the third problem. The case of a source on-axis, at a finite distance, represents the simplest case in which there is geometrical vignetting for double reflection. If the source is on-axis, at infinite distance (with $L_2 \geq L_1$), all rays undergo two reflections and reach the focus of the mirror. If the distance is finite, not all photons reflected by the primary segment also hit the secondary: for instance, photons reflected at $z \approx+L_1$ can miss being reflected by the secondary, so they are prevented from being focused, as depicted in Fig.~\ref{fig:mirror_section}. We can quantify this geometric effect by means of the {\it vignetting factor} $V$, which is defined to be the ratio of the number of photons reflected twice to the number of photons reflected by the primary segment (assuming, ideally, that the mirror reflectivity is 1). In this configuration, if the mirror profile is a double cone, we can compute $V$ using the simple formula (see Appendix~\ref{App_vign} for a derivation)
\begin{equation}
	V \simeq \frac{L_2\,\alpha_2}{L_1\,\alpha_1},
	\label{eq:vign}
\end{equation}
valid if $L_2\alpha_2 < L_1\alpha_1$, otherwise $V =1$. If either $\alpha_1~<~0$ or $\alpha_2~<~0$, then $V =0$. In Sect.~\ref{EffArea}, we see that this result can even be generalized to the case of a source off-axis. 

If we consider that in this case $\alpha_1~=~\alpha_0+\delta$ and $\alpha_2~=~\alpha_0-\delta$, an alternative expression for Eq.~(\ref{eq:vign}) is
\begin{equation}
	V \simeq \frac{L_2\, (D-4f)}{L_1\,(D+4f)},
	\label{eq:vign2}
\end{equation}
where we have used the definitions of $\alpha_0$ and $\delta$. Using Eq.~(\ref{eq:f_num}) and normalizing the $L$'s to the mirror diameter, we can also write Eq.~(\ref{eq:vign}) as 
\begin{equation}
	V \simeq \frac{L'_2\,(1-8\delta\,f\#)}{L'_1\,(1+8\delta\,f\#)}.
	\label{eq:vign_fnum}
\end{equation}
It is interesting to note that a formula similar to Eq.~(\ref{eq:vign2}) for the effective area was empirically found by Van Speybroeck and Chase (\cite{VanSpeybroeck}), but with a 2.5$f$ term instead of 4$f$. 

Rigorously, Eqs.~(\ref{eq:vign}) to~(\ref{eq:vign_fnum}) hold only for a double cone mirror, but they are expected to be applicable also to Wolter-I mirrors of a sufficiently large $f\#$. To determine the approximation that can be achieved, we computed by ray-tracing the {\it exact} $V$ factor of a Wolter-I mirror as a function of $f\#$, in the case of $L'_1 = L'_2 = 0.5$, for fixed $\delta$ values. In Fig.~\ref{fig:fin_dist}, the results are compared with the findings of Eq.~(\ref{eq:vign_fnum}). The vignetting for the double cone is independent of $L'$, whereas for the Wolter's it depends markedly on $L'$ only for small $f\#$. 

Inspection of Fig.~\ref{fig:fin_dist} shows that
\begin{itemize}
	\item{for a fixed $\delta$ and not too large $f\#$, $V$ is {\it larger} for a Wolter-I mirror than for the corresponding double cone; therefore, the double cone approximation returns, in general, pessimistic predictions of the effective area.}
	\item{In double cone approximation, $V<1$ for all $f\#$, i.e., there are always lost photons for double reflection. With a Wolter-I profile, $V$ is 1 when $f\#$ becomes sufficiently small.}
	\item{For a fixed $\delta$, the predictions of Eq.~(\ref{eq:vign}) approach the exact calculations as $f\#$ is increased. For large $f\#$ and large $\delta$, where $V\simeq 0$, the $V$ factor of Wolter becomes {\it slightly smaller} than that of the double cone.}
\end{itemize}

\begin{figure}
	\resizebox{\hsize}{!}{\includegraphics{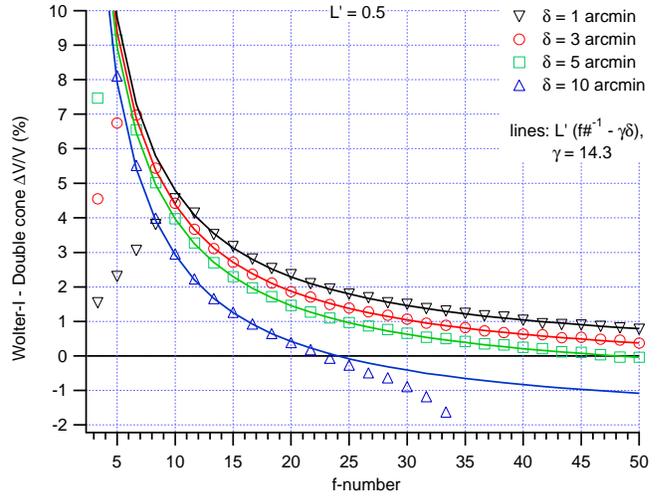}}
	\caption{Percent accuracy of the double cone approximation in the estimation of the vignetting factor for a source {\it on-axis} at finite distance, as derived from the curves of Fig.~\ref{fig:fin_dist} (symbols). Lines are traced using the empirical Eq.~(\ref{eq:F_var}) with $L'~=~0.5$.}
\label{fig:error_eval}
\end{figure}

From the curves of Fig.~\ref{fig:fin_dist}, we computed $\frac{\Delta V}{V}$, the error in the vignetting factor when we apply Eq.~(\ref{eq:vign}) to a Wolter-I profile, as in the definition of Eq.~(\ref{eq:max_diam}). This quantity is plotted in Fig.~\ref{fig:error_eval} for the case $L' = 0.5$. It can be seen that, in the cases interesting for us, i.e., for sufficiently large $f\#$ and not too large values of $\delta$, $\frac{\Delta V}{V}$ is positive and of the order of a few percent. We repeated this exercise for several values of $L'$ in the interval $0.25 - 1$ to investigate the dependence of $\frac{\Delta V}{V}$ on this parameter, and it turned out that this ratio, if positive and if $V(\mbox{Wolter}) < 1$, can be approximated very well by the empirical formula (see Fig.~\ref{fig:error_eval} and Table~\ref{tab:DC_toler})
\begin{equation}
	\frac{\Delta V}{V}\simeq L'\left(\frac{1}{f\#}-\gamma\delta\right),
	\label{eq:F_var}
\end{equation}
where $\gamma \approx 14.3$ is with very good approximation a constant in the explored range of $L'$ values. This equation has the same kind of dependence as Eq.~(\ref{eq:angle_var1}) for the slope variation along the profile of the mirror and Eq.~(\ref{eq:max_diam}) for the area of the primary segment. 

For small $f\#$ values, Eq.~(\ref{eq:F_var}) is not obeyed because of the saturation of $V$(Wolter) to 1, so the expression of $\frac{\Delta V}{V}$ should be interpreted as an upper limit. Finally, for very large $\delta$ the error deviates from Eq.~(\ref{eq:F_var}) as it becomes negative, but this occurs only when $V \rightarrow 0$, so its weight in determining the effective area is expected to be negligible.

We can therefore conclude that the error introduced by the double cone approximation, regarding the effective area, is definitively smaller than $L'/f\#$. In other words, the double cone approximation is valid when $L \ll f$, a condition fulfilled in almost all practical cases. Within the limits of this approximation, we derive in the next section the effective area for a source on- and off-axis.

\section{The on- and off-axis effective area of a Wolter-I mirror}
\label{EffArea}
\subsection{On-axis source}\label{onaxis}
We assume that we can approximate the Wolter-I profile with a double cone profile, by adopting the tolerances estimated in Sect.~\ref{DC_WI}. In the following, we adopt the convention to denote with $A_D(\lambda, \theta)$ the {\it effective} area of the mirror at the photon wavelength $\lambda$, for a source off-axis by $\theta$, at a distance $D$ (finite or infinite). When referring to the {\it geometric} area, we omit $\lambda$ and use the notation $A_D(\theta)$. Firstly, we assume $L_1 = L_2 = L$. The geometrical, collecting area of the primary segment, as seen by a source on-axis at ``infinite" (i.e., astronomical) distance is 
\begin{equation}
	A_{1,\infty}(0) = \pi(R_{\mathrm M}^2-R_0^2) \simeq 2\pi R_0\,(R_{\mathrm M}-R_0) \simeq 2\pi R_0 L\,\alpha_0,
\label{eq:Ag_inf_on}
\end{equation}
where we use the approximation $R_{\mathrm M} - R_0 \simeq L \alpha_0$. In this case, all reflected rays also undergo an identical reflection on the secondary segment (Fig.~\ref{fig:mirror_section}), therefore $A_{\infty}(0)=A_{1,\infty}(0)$. We now denote by $r_{\lambda}(\alpha)$ the reflectivity of the mirror at the photon wavelength $\lambda$, for a generic incidence angle $\alpha$. The form of this function depends on the coating structure: for a single layer coating, which operates in total external reflection, it slowly decreases up to the critical angle for $\lambda$, followed by a sudden cutoff. If a multilayer coating is used, $r_{\lambda}(\alpha)$ is a more complicated function and can be computed using one of the standard methods (e.g., Parrat~\cite{Parrat}; Abel\`es~\cite{Abeles}), by also including the effect of roughness using, e.g., the N\'evot-Croce~(\cite{NevotCroce}) approach.

Since photons are reflected twice at the same angle $\alpha_0$, we multiply the geometrical area by the squared reflectivity to obtain the {\it effective} area at the photon wavelength $\lambda$
\begin{equation}
	A_{\infty}(\lambda, 0) = A_{\infty}(0)\cdot r^2_{\lambda}(\alpha_0) =2\pi R_0 L\,\alpha_0\cdot r^2_{\lambda}(\alpha_0).
\label{eq:Ae_inf_on}
\end{equation}
This is a well known result. We now keep the source on-axis, but at a finite distance $D$ and assume more generally that $L_1 \neq L_2$ (with $L_1 = L_2$ as a particular case). As already discussed in Sect.~\ref{DC_WI}, all mirror sectors see the source off-axis by the same angle $\delta = R_0/D$. The effective area of the {\it primary} segment thereby becomes
\begin{equation}
	A_{1,D}(\lambda, 0) = 2\pi R_0 L_1\alpha_1\cdot r_{\lambda}(\alpha_1),
\label{eq:Apar_fin_on}
\end{equation}
where $\alpha_1 = \alpha_0+\delta$. The effective area for a source at finite distance is then obtained from Eq.~(\ref{eq:Apar_fin_on}), times the vignetting factor of Eq.~(\ref{eq:vign}), times $r_{\lambda}(\alpha_2)$, the reflectivity of the secondary mirror segment,
\begin{equation}
	A_D(\lambda, 0) = 2\pi R_0 L_1\alpha_1 V \cdot r_{\lambda}(\alpha_1) r_{\lambda}(\alpha_2),
\label{eq:Aeff_fin_on}
\end{equation}
where $\alpha_2 = \alpha_0-\delta$. {\it If} $V<1$, substitution of Eq.~(\ref{eq:vign}) yields 
\begin{eqnarray}
	A_D(\lambda, 0) =2\pi R_0 \cdot L_2\alpha_2 \cdot r_{\lambda}(\alpha_1)\,r_{\lambda}(\alpha_2),& \mbox{if} & L_2\,\alpha_2 < L_1\,\alpha_1.
\label{eq:case1}
\end{eqnarray}
As one might expect, the geometric area is that of the secondary segment, projected onto the wavefront after the primary reflection. We note that by setting $\delta = 0$ we retrieve the on-axis result, in Eq.~(\ref{eq:Ae_inf_on}). On the other hand, if $L_2\alpha_2 >L_1\alpha_1$, all the primary segment is effective in the double reflection, so $V =1$. This may occur with a divergent source on-axis, if $L_2 \gg L_1$. In the absence of a geometrical vignetting, the effective area becomes 
\begin{eqnarray}
	A_D(\lambda, 0) = 2\pi R_0 \cdot L_1\alpha_1 \cdot r_{\lambda}(\alpha_1)\,r_{\lambda}(\alpha_2) & \mbox{if} & L_2\,\alpha_2 > L_1\,\alpha_1.
\label{eq:case2}
\end{eqnarray}
Comparison of Eqs.~(\ref{eq:case1}) and~(\ref{eq:case2}) indicates that the effective area can be written as 
\begin{equation}
	A_D(\lambda, 0) =2\pi R_0 \,\min(L_1\alpha_1, L_2\alpha_2) \,r_{\lambda}(\alpha_1)\,r_{\lambda}(\alpha_2),
\label{eq:case_gen}
\end{equation}
which represents the general expression for the effective area seen by a source on-axis.

\subsection{Off-axis source: integral formula} 
\label{offaxis}
\begin{figure}
	\resizebox{\hsize}{!}{\includegraphics{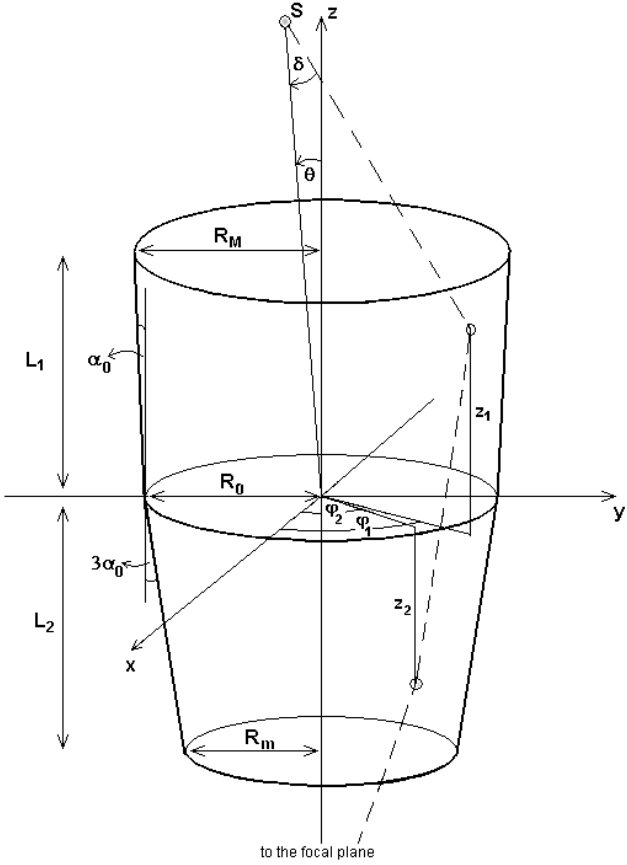}}
	\caption{A grazing-incidence double cone mirror, illuminated by an off-axis source: the dashed line is a ray path. The optical axis lies along the $z$ direction and the $x$ axis is chosen for the source $S$, at a distance $D = R_0/\delta$, to lie in the $xz$ plane. The source direction forms an angle $\theta$ with the $z$ axis. The azimuthal positions of the reflection points are located along with the $\varphi_1$ and $\varphi_2$ polar angles.}
\label{fig:mirror_scheme}
\end{figure}

We can now compute the effective area for a source off-axis. Because of the axial symmetry of the mirror, we choose the $x$ axis so that the source lies in the $xz$ plane, on the side of the positive $x$ axis (refer to Fig.~\ref{fig:mirror_scheme}). We define $\theta > 0$ to be the angle between $z$ and the source direction. 

With respect to the on-axis case, there are some additional difficulties. The ray no longer lies within a meridional plane of the mirror, so the polar angles of the impact positions on the primary and secondary mirror segment, $\varphi_1$ and $\varphi_2$, differ in general, and the incidence angles $\alpha_1$ and $\alpha_2$ also vary with them. 

Nevertheless, since the maximum distance of the two impact points is $\sim L_1+L_2 \approx 2L$, the off-plane linear displacement is $2\theta L$ at most: therefore, $|\varphi_2-\varphi_1|\lesssim 4\theta L'$, which is in general negligible with respect to $\varphi_1$ and $\varphi_2$ themselves. This result can also be derived more rigorously (see Appendix~\ref{App_angl}). 

In the limit of small $\theta$, we can henceforth assume with good approximation $\varphi_1 \approx \varphi_2$: for this reason, in the remainder of this section we denote with $\varphi$ the nearly-common value of the two angles. In Appendix~\ref{App_angl}, we also derived other important results: 
\begin{enumerate}
	\item{analytical expressions for $\alpha_1(\varphi)$ and $\alpha_2(\varphi)$ can be found with some algebra. For a small angle $\theta$, these functions reduce to the simple expressions
\begin{eqnarray}
	\alpha_1(\varphi) & \simeq & \alpha_0+\delta-\theta\cos\varphi \label{eq:angles1}\\
	\alpha_2(\varphi) & \simeq & \alpha_0-\delta+\theta\cos\varphi \label{eq:angles2},
\end{eqnarray}
which generalize the expressions for the incidence angles of Sect.~\ref{onaxis}, to the case of an off-axis source;}
	\item{always in the limit of small $\alpha_0$, $\delta$, $\theta$, the vignetting for double reflection (Sect.~\ref{vignetting}) can be calculated using a generalization of Eq.~(\ref{eq:vign}), 
\begin{equation}
	V(\varphi) \simeq \frac{L_2\, \alpha_2(\varphi)}{L_1\,\alpha_1(\varphi)}.
	\label{eq:vign_gen}
\end{equation}
Here the $V$ factor varies with $\varphi$ because it depends on $\alpha_1$ (Eq.~(\ref{eq:angles1})) and $\alpha_2$ (Eq.~(\ref{eq:angles2})). As in Eq.~(\ref{eq:vign}), Eq.~(\ref{eq:vign_gen}) is valid only if $0~<~V(\varphi)~<~1$. If $V$ exceeds 1 at some $\varphi^*$, then all the primary segment sector is effective and $V(\varphi^*)=1$. If either $\alpha_1$ or $\alpha_2$ is negative at some $\varphi^*$, no reflection occurs on the optical side of the mirror and $V(\varphi^*)=0$.}
\end{enumerate}

We now draw our attention to a small mirror sector between $\varphi$ and $\varphi+\Delta\varphi$, when $\Delta\varphi \rightarrow 0$. Equations~(\ref{eq:angles1}) and~(\ref{eq:angles2}) can then be used to compute the local incidence angles, and Eq.~(\ref{eq:vign_gen}) to compute the local vignetting factor. We assume, initially, that $\alpha_1 >0$ and $\alpha_2 >0$; therefore $V>0$. If $V(\varphi) <1$, we can repeat the passages in Sect.~\ref{onaxis} for a mirror sector at $\varphi$ according to Eq.~(\ref{eq:case1}). Otherwise $V(\varphi)=1$, all the primary segment sector is effective, so we can write the sector area according to Eq.~(\ref{eq:case2}). Finally, we obtain for the effective area of the infinitesimal sector, 
\begin{equation}
	\Delta A_D(\lambda, \theta, \varphi) =R_0 \,(L\alpha)_{\mathrm{min}} \,r_{\lambda}(\alpha_1)\,r_{\lambda}(\alpha_2) \,\Delta\varphi,
\label{eq:ea_sector}
\end{equation}
where $(L\alpha)_{\mathrm{min}}= \min[L_1 \alpha_1(\varphi), L_2\alpha_2(\varphi)]$. In Eq.~(\ref{eq:ea_sector}), we omitted the explicit dependence of $\alpha_1$ and $\alpha_2$ on $\varphi$: in the following, we adopt the same convention to simplify the notation. 

The expression for $(L\alpha)_{\mathrm{min}}$ is correct as long as the incidence angles are non-negative. If either $\alpha_1$ or $\alpha_2$ become negative at some $\varphi$, no contribution to the effective area can be given by that sector, so $(L\alpha)_{\mathrm{min}}$ is zero at that $\varphi$. We can then write without restrictions that 
\begin{equation}
	(L\alpha)_{\mathrm{min}} = \max\left\{0, \min[L_1 \alpha_1(\varphi), L_2 \alpha_2(\varphi)]\right\}.
	\label{eq:Lalpha_min}
\end{equation}
The effective area of the entire mirror is then given by integration over $\varphi$
\begin{equation}
	A_D(\lambda, \theta) = 2R_0 \int_0^{\pi}\!\!(L\alpha)_{\mathrm{min}} \,r_{\lambda}(\alpha_1)\,r_{\lambda}(\alpha_2) \,\mbox{d}\varphi,
\label{eq:Aeff_fin_offaxis}
\end{equation}
where the factor of 2 comes from the symmetry of the mirror with respect to the off-axis plane. In the particular case $L_1 = L_2$, Eq.~(\ref{eq:Aeff_fin_offaxis}) becomes
\begin{equation}
	A_{D}(\lambda, \theta) = 2R_0 L\int_0^{\pi}\!\!\alpha_{\mathrm{min}}\,r_{\lambda}(\alpha_1)\,r_{\lambda}(\alpha_2)\, \mbox{d}\varphi,
\label{eq:Aeff_fin_offaxis0}
\end{equation}
where $\alpha_{\mathrm{min}} = \min[\alpha_1(\varphi), \alpha_2(\varphi)]$ if positive, and zero otherwise. Using Eqs.~(\ref{eq:angles1}) and~(\ref{eq:angles2}), we can conveniently write $\alpha_{\mathrm{min}}$ in the compact form
\begin{equation}
	\alpha_{\mathrm{min}} = \max(0,\,\alpha_0-|\delta-\theta\cos\varphi|),
\label{eq:theta_min}
\end{equation}
which is valid also for $\delta > \alpha_0$. Finally, for a source at infinity there is additional symmetry with respect to the $y$ axis, and, if $\theta~<~\alpha_0$, we can write Eq.~(\ref{eq:Aeff_fin_offaxis0}) in an even simpler form,
\begin{equation}
	A_{\infty}(\lambda, \theta) = 4R_0 L\int_0^{\pi/2}\!\!\!(\alpha_0-\theta\cos\varphi)\,r_{\lambda}(\alpha_1)\,r_{\lambda}(\alpha_2)\, \mbox{d}\varphi.
\label{eq:Aeff_fin_infty}
\end{equation}

\section{Applications to the geometric area}
\label{Geometric}
An upper limit to the effective area $A_{D}(\lambda, \theta)$ is represented by the geometric area, $A_{D}(\theta)$, obtained by simply setting the mirror reflectivity to 1, for all $\lambda$ and $\alpha$. In the most common case, $L_1~=~L_2$, the expression of the geometric area is
\begin{equation}
	A_{D}(\theta) = 2R_0 L\int_0^{\pi}\!\alpha_{\mathrm{min}}\, \mbox{d}\varphi,
\label{eq:Ageom_fin_off}
\end{equation}
where $\alpha_{\mathrm{min}}$ is given by Eq.~(\ref{eq:theta_min}). In this section we solve this integral explicitly and provide analytical expressions for $A_D(\theta)$.

\subsection{Source at infinity}
For a source at infinity with $\theta < \alpha_0$, it is more convenient to use Eq.~(\ref{eq:Aeff_fin_infty}) with $r_{\lambda}(\alpha) = 1$
\begin{equation}
	A_{\infty}(\theta) = 4R_0 L\int_{0}^{\pi/2}\!\!\!(\alpha_0- \theta\cos\varphi) \,\mbox{d}\varphi,
\label{eq:Ageom_SC1}
\end{equation}
which can be immediately solved
\begin{equation}
	A_{\infty}(\theta) = 2\pi R_0 L\alpha_0- 4R_0 L\theta.
\label{eq:Ageom_SC2}
\end{equation}
Recalling Eq.~(\ref{eq:Ae_inf_on}) for the on-axis area $A_{\infty}(0)$, we obtain 
\begin{equation}
	A_{\infty}(\theta) \stackrel{\theta < \alpha_0}{=} A_{\infty}(0)\, \left(1- \frac{2\theta}{\pi\alpha_0}\right),
\label{eq:Ageom_SC3}
\end{equation}
which is exactly Eq.~(\ref{eq:SC_formula}) found by Van Speybroeck and Chase~(\cite{VanSpeybroeck}), after approximating $\pi \simeq 3$. 

For off-axis angles larger than $\alpha_0$, Eq.~(\ref{eq:Ageom_SC3}) is no longer valid. An extension of the curve $A_{\infty}(\theta)$ for $\theta > \alpha_0$ can be obtained from Eq.~(\ref{eq:Ageom_SC1}), after setting the integrand to 0 when $\cos\varphi > \alpha_0/\theta$, according to Eq.~(\ref{eq:theta_min}). The result is a non-linear function of the $\theta/\alpha_0$ ratio,
\begin{equation}
	A_{\infty}(\theta) \stackrel{\theta > \alpha_0}{=} A_{\infty}(0)\, \left[1-\frac{2}{\pi}\left(\frac{\theta}{\alpha_0}-\sqrt{\frac{\theta^2}{\alpha_0^2}-1}+\arccos\frac{\alpha_0}{\theta}\right)\right].
\label{eq:Ageom_SC4}
\end{equation}
which is identical to Eq.~(\ref{eq:Ageom_SC3}) for $\theta = \alpha_0$; this is correct because the geometrical vignetting must be a continuous function of $\theta$. We note that, for sufficiently large $\theta$, Eq.~(\ref{eq:Ageom_SC4}) can be approximated well by $A_{\infty}(0)\,\alpha_0/(\pi\theta)$. 

A complete vignetting curve for $\delta = 0$ is shown in Fig.~\ref{fig:geom_vignetting} (solid line). The predicted deviation from linearity is verified in Sect.~\ref{Comp} by means of an accurate ray-tracing routine. 

\subsection{Source at finite distance}
We now consider the variation in the geometric area, for a source at a finite distance ($\delta >0$). In this case, the integration depends on whether $\delta<\alpha_0/2$ or not. 

We consider firstly the case $\delta<\alpha_0/2$. We then assume initially $\theta < \delta$. With these conditions, $\theta\cos\varphi < \delta < \alpha_0-\delta$ for all~$\varphi$, therefore Eq.~(\ref{eq:theta_min}) becomes
\begin{equation}
	\alpha_{\mathrm{min}}(\varphi) = \alpha_0-\delta+\theta\cos\varphi.
\label{eq:Ageom_D1}
\end{equation}
Substituting this expression into Eq.~(\ref{eq:Ageom_fin_off}) and solving, we derive the area 
\begin{equation}
	A_{D}(\theta) = 2\pi R_0 L\,(\alpha_0-\delta),
\label{eq:Ageom_D3}
\end{equation}
that is, $A_{D}(\theta)= A_{D}(0)$ (see Eq.~(\ref{eq:case1})). In other words, the mirror geometric area is constant as far as $\theta<\delta<\alpha_0/2$. This is often observed in optics calibrations at on-ground facilities (see, e.g., Gondoin~et~al.~\cite{Gondoin2}). 

We now increase $\theta$ beyond $\delta$. Since $\delta < \alpha_0-\delta$ by hypothesis, we can consider the case $\delta <\theta< \alpha_0-\delta < \alpha_0+\delta$. We are therefore allowed to write Eq.~(\ref{eq:Ageom_fin_off}) as
\begin{eqnarray}
	A_{D}(\theta) &= & 2R_0 L\int_0^{\arccos\frac{\delta}{\theta}}\!\!\!\!\!(\alpha_0+\delta-\theta\cos\varphi)\, \mbox{d}\varphi + \nonumber\\
	& + & 2R_0 L\int_{\arccos\frac{\delta}{\theta}}^{\pi}\!\!\!\!\!(\alpha_0-\delta+\theta\cos\varphi)\, \mbox{d}\varphi,
\label{eq:Ageom_D4}
\end{eqnarray}
where the integrands are always positive. This yields
\begin{equation}
	A_{D}(\theta)= A_{\infty}(0) \left[1 - \frac{2\delta}{\pi\alpha_0}\left(\arcsin\frac{\delta}{\theta}+ \sqrt{\frac{\theta^2}{\delta^2}-1}\,\right)\right].
\label{eq:Ageom_D5}
\end{equation}
We note that for $\delta \rightarrow 0$, Eq.~(\ref{eq:Ageom_D5}) reduces to Eq.~(\ref{eq:Ageom_SC3}), and for $\theta \rightarrow \delta$ to Eq.~(\ref{eq:Ageom_D3}), as expected.

We now suppose $\alpha_0-\delta <\theta< \alpha_0+\delta$. In this case, the integration returns some more terms
\begin{eqnarray}
	A_{D}(\theta) & =& A_{\infty}(0) \left\{1 - \frac{2\delta}{\pi\alpha_0}\left(\arcsin\frac{\delta}{\theta}+\sqrt{\frac{\theta^2}{\delta^2}-1}\,\right)\right.+\nonumber\\
	&+&\left. \frac{1}{\pi\alpha_0}\left[\sqrt{\theta^2-(\alpha_0-\delta)^2} -(\alpha_0-\delta)\arccos\frac{\alpha_0-\delta}{\theta}\right]\right\}.
\label{eq:Ageom_D6}
\end{eqnarray}
This equation, as expected, returns the same result as Eq.~(\ref{eq:Ageom_D5}) at $\theta=\alpha_0-\delta$, and is valid for off-axis angles in the interval $[\alpha_0-\delta, \alpha_0+\delta]$. Therefore, for $\delta \rightarrow 0$ it reduces to Eq.~(\ref{eq:Ageom_SC3}) and~(\ref{eq:Ageom_SC4}) for the single point $\theta = \alpha_0$. 

\begin{figure}
	\resizebox{\hsize}{!}{\includegraphics{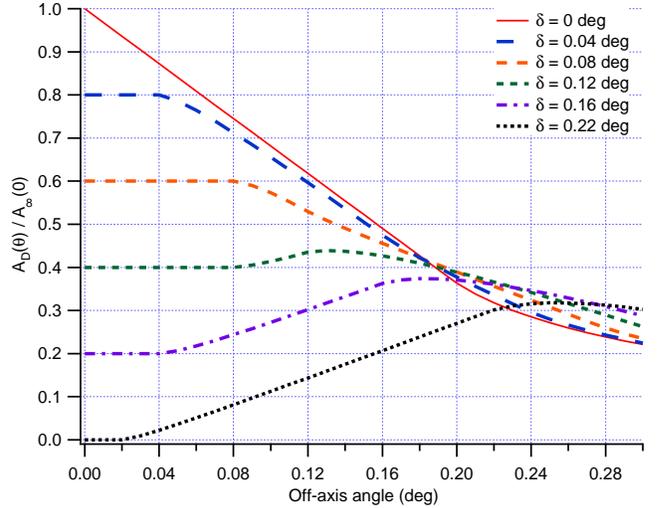}}
	\caption{Normalized geometric area, $A_{D}(\theta)/A_{\infty}(0)$, for a double cone mirror with a $\alpha_0 = 0.2$~deg, as a function of the off-axis angle, for different $\delta$ angles due to the finite distance of the source. The curves are traced using the analytical formulae reported in Sect.~\ref{Geometric}.}
\label{fig:geom_vignetting}
\end{figure}

Now, consider the case $\delta > \alpha_0/2$, unlike we have hitherto assumed, but still that $\delta <\alpha_0$. This time $\alpha_0-\delta <\delta$: the condition $\theta<\delta$ is insufficient for avoiding negative values of $\alpha_{\mathrm{min}}$ in Eq.~(\ref{eq:Ageom_D1}), therefore Eq.~(\ref{eq:Ageom_D3}) is valid only if $\theta<\alpha_0-\delta$. Beyond this limit and up to $\theta = \delta$, Eq.~(\ref{eq:Ageom_D5}) should be replaced by 
\begin{equation}
	A_{D}(\theta) = \frac{A_D(0)}{\pi} \left(\arccos\frac{\delta- \alpha_0}{\theta}+ \sqrt{\frac{\theta^2}{(\alpha_0-\delta)^2}-1}\,\right),
\label{eq:Ageom_D7}
\end{equation}
which, unexpectedly, is an {\it increasing} function of $\theta$ in the interval $\alpha_0-\delta<\theta<\delta$. When $\delta <\theta <\delta+\alpha_0$, the geometric area again follows Eq.~(\ref{eq:Ageom_D6}), which for $\delta >\alpha_0/2$ exhibits a {\it maximum} at
\begin{equation}
	\theta_{\mathrm{max}} = \sqrt{\delta^2+\frac{2}{3}\alpha_0\delta-\frac{1}{3}\alpha_0^2}:
	\label{eq:theta_max}
\end{equation}
such a maximum is not present if $\delta < \alpha_0/2$.

In a similar fashion, one can easily compute the total geometric vignetting for $\theta>\alpha_0+\delta$, even if the resulting expression would be too long to report here, and it is also possible to derive similar expressions for the unusual case $\delta > \alpha_0$. To provide the reader with a qualitative visualization of the overall trend of the expressions we just derived, we traced in Fig.~\ref{fig:geom_vignetting} some curves of total vignetting as a function of $\theta$, for $\alpha_0 = 0.2$~deg and different values of $\delta$. To do this, we used Eqs.~(\ref{eq:Ageom_SC3}) and~(\ref{eq:Ageom_SC4}) and Eqs.~(\ref{eq:Ageom_D3}) to (\ref{eq:Ageom_D7}) in the respective intervals of validity, and also the expression for $\theta > \alpha_0+\delta$, which is not reported here. 

To summarize the results presented in this section, from Fig.~\ref{fig:geom_vignetting} we can draw some conclusions:
\begin{itemize}
	\item{if $\delta =0$, the geometric area decreases linearly up to $\theta = \alpha_0$, then approaches zero as $\theta$ is increased beyond $\alpha_0$;}
	\item{if $0< \delta < \alpha_0/2$, the geometric area remains constant for $\theta< \delta$, then decreases monotonically to zero;}
	\item{if $\alpha_0/2<\delta<\alpha_0 $, the geometric area remains constant for $\theta~<~\alpha_0-\delta$, then increases, reaches a maximum and eventually decreases to zero;}
	\item{if $\delta > \alpha_0$, the geometric area is zero for $\theta < \delta-\alpha_0$, then increases, reaches a maximum and eventually decreases to zero.}
\end{itemize}
	
\section{Validation with ray-tracing results}
\label{Comp}
We verify the formulae derived in previous sections by means of a comparison with the results of a ray-tracing routine. To make the comparison easier, we consider the case of a single Wolter-I mirror. A first point to be checked is the geometric area, which was extensively analyzed from the analytical viewpoint in Sect.~\ref{Geometric}. In particular, the curves of Fig.~\ref{fig:geom_vignetting} are easily verifiable along with a ray-tracing routine, without any assumptions about the reflective coating. The comparison is interesting especially when testing the analytical formulae in the non-linear regions, which are usually not exploited in X-ray optics.

\begin{figure}
	\resizebox{\hsize}{!}{\includegraphics{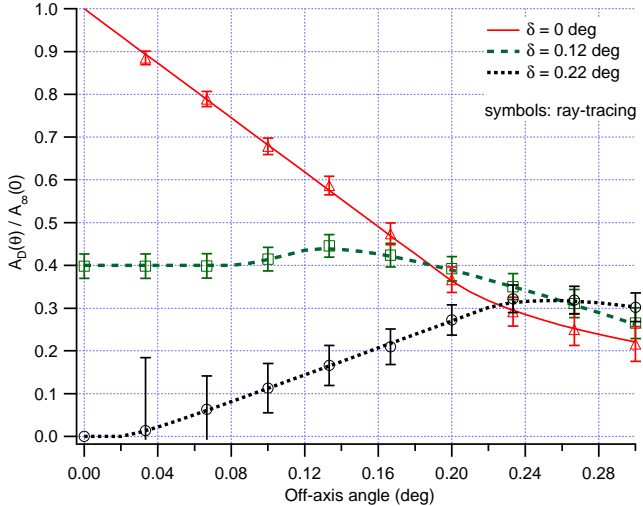}}
	\caption{Comparison between some analytical curves of Fig.~\ref{fig:geom_vignetting} (lines), and the results of an accurate ray-tracing (symbols) for a Wolter-I mirror with the same $\alpha_0$ (0.2 deg) and $L'$~=~0.5. To avoid confusion in the figure, not all curves were represented. Note the very good matching of the two methods, within the error bars of the ray-tracing.}
\label{fig:vign_comp}
\end{figure}

We displayed in Fig.~\ref{fig:vign_comp} the comparison between 3 representative analytical curves of normalized geometric area (from Fig.~\ref{fig:geom_vignetting}) and the findings of a ray-tracing run on a Wolter-I mirror with the same value of $\alpha_0$ (0.2~deg), corresponding to $f\# \approx~36$. For the ray-tracing simulation, we adopted the reasonable value $L'$~=~0.5, whereas the analytical curves are largely independent of this parameter, as long as the double cone approximation is applicable. More exactly, from the discussion of Sect.~\ref{DC_WI} the approximation that we introduce is not larger than $L'/f\# = 1.5\%$. As can be noted, the agreement between the two methods is complete, also for large off-axis angles where the vignetting ceases to be linear, and even for $\delta > \alpha_0$. On the other hand, the analytical computation, with a nearly-continuous $\theta$ sampling, corresponded to a few lines of IDL code, whilst the ray-tracing routine is a complex program, which requires the acquisition of more than $10^4$ photons to reach a statistical error of a few percent for each point.
 
\begin{figure}
	\resizebox{\hsize}{!}{\includegraphics{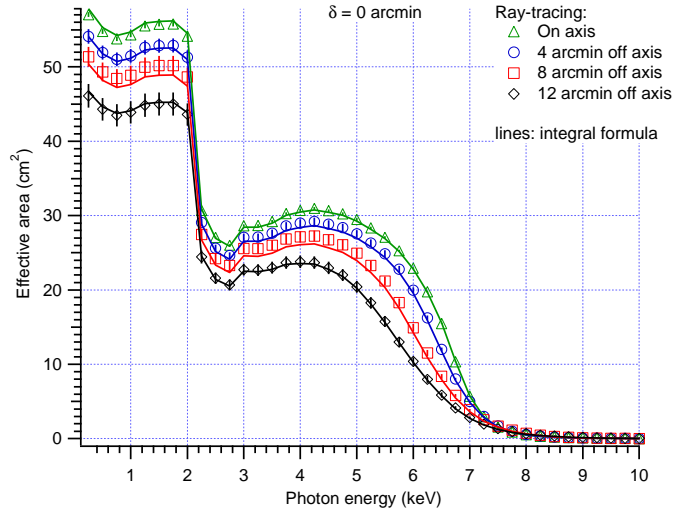}}
	\caption{Comparison of effective areas of the largest Wolter-I mirror of Newton-XMM, as computed from ray-tracing (symbols) and from Eq.~(\ref{eq:Aeff_fin_offaxis0}) (lines). The source is supposed to be at infinity ($\delta =0$). The agreement is very good to within a few percent.}
\label{fig:XMM_mir}
\end{figure}

We now wish to compare the predictions of Eq.~(\ref{eq:Aeff_fin_offaxis0}) for the effective area, as a function of the photon energy, with the results of the ray-tracing. As a first case, we check the results for the largest mirror shell of a Newton-XMM optical module (Gondoin~et~al.~\cite{Gondoin2}), with $f$~=~7.5~m, $R_0$~=~346.2~mm ($\alpha_0$~=~0.66~deg), and $L$~=~300~mm. The errors in the primary mirror segment area and the incidence angle variation along the profile introduced by the double cone approximation are less than 1\%, and the error in the vignetting for double reflection is 4\% {\it at most} (see Table~\ref{tab:DC_toler}), so Eq.~(\ref{eq:Aeff_fin_offaxis0}) can be reliably applied to compute the effective area of the mirror, both on- and off-axis. 

The comparison between the results obtained from a ray-tracing (symbols) and from the application of Eq.~(\ref{eq:Aeff_fin_offaxis0}) (lines) is shown in Fig.~\ref{fig:XMM_mir}, for a source at infinity and 4 different values of the off-axis angle. For both computational methods, the reflectivity of the Gold coating was computed by assuming the same surface roughness rms of $\sigma$~=~4~\AA, regardless of the photon energy and the incidence angle. This is not completely correct, because the spectral window of the roughness power spectrum effective for X-ray scattering changes with $\lambda$, $\alpha_1$, $\alpha_2$, and the size of the region over which the image is integrated. A variable $\sigma$, computed from the power spectrum of roughness in a variable frequency range, should be adopted (Spiga~et~al.~\cite{Spiga2009}). Nevertheless, this has no relevance to the present comparison and we retain $\sigma$ as a constant. 

As can be noted from Fig.~\ref{fig:XMM_mir}, the comparison provides a good agreement between the two methods, at all considered energies and off-axis angles. The analytical method underestimates the ray-tracing findings by only 3.7\% at most, as foreseen, an amount close to the statistical error for the ray-tracing. This confirms the correctness of the analytical formula (Eq.(\ref{eq:Aeff_fin_offaxis0})) for Wolter-I mirrors within the approximation limits stated in Sect.~\ref{DC_WI}. 

Finally, as an application to the hard X-ray band ($>$ 10 keV), we consider the case of a long focal length ($f = 20$ m) X-ray mirror with a wideband multilayer coating. Because of the complex dependence of multilayer reflectivity on the photon energy and the incidence angles, this example places the analytical method to the test, because any departure of the incidence angles from the true values would result in a displacement of the reflectance peaks. As simulation parameters, we assumed $R_0~=~296.2$ mm, and $L_1 = L_2 = 300$ mm, $\alpha_0$ = 0.106 deg. The geometrical area for a source on-axis, at infinity, would be 5.17~cm$^2$. The ratio expressing the departure of the double cone from the Wolter's, $L'/f\#$, is only 1.5\%. The multilayer coating is supposed to consist of 200 pairs of Pt/C layers. The layer thickness decreases from the coating surface towards the substrate, according to the well-known power-law for the d-spacing -- i.e., the sum of the thicknesses of two adjacent layers -- $d_j = a(b+j)^{-c}$ (Joensen \cite{Joensen}), with $j =$1, 2, \ldots, 200 and $a$, $b$, $c$, parameters with values depending on the desired reflectivity. In the present example we adopted $a = 115.5~\AA$, $b = 0.9$, and $c = 0.27$, for a constant thickness ratio of Pt to the d-spacing, $\Gamma = 0.35$. The outermost layer is Pt. Finally, the surface roughness of the mirror is assumed to have the constant value $\sigma = 4~\AA$.

\begin{figure}
	\resizebox{\hsize}{!}{\includegraphics{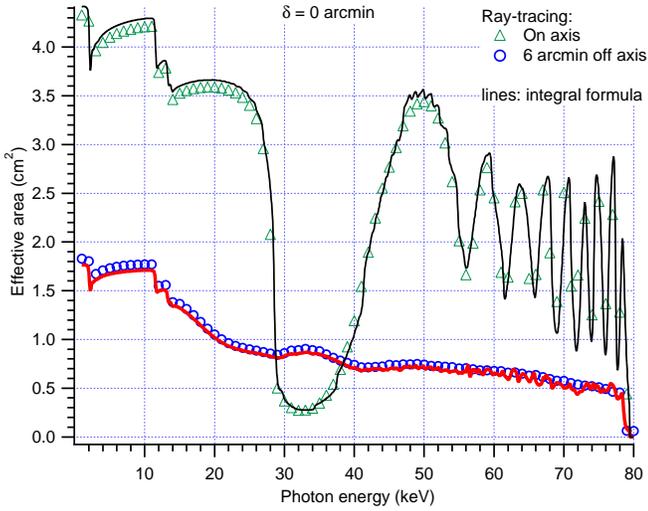}}
	\caption{Comparison of effective areas of a multilayer-coated Wolter-I mirror, as computed from a ray-tracing (symbols) and from Eq.~(\ref{eq:Aeff_fin_offaxis0}) (lines). The source is supposed to be at infinity ($\delta =0$). The error bars of the ray-tracing outputs are not shown. The accord between the curves is within a few percent on-axis (triangles) and 6~arcmin off-axis (circles).}
\label{fig:comp_infty}
\end{figure}
\begin{figure}
	\resizebox{\hsize}{!}{\includegraphics{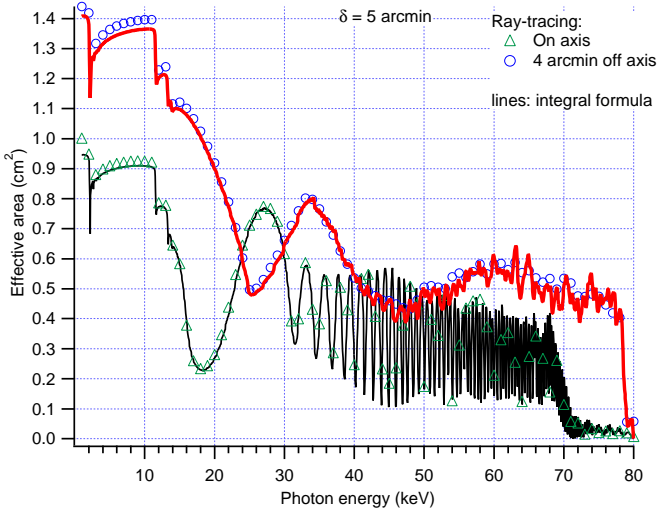}}
	\caption{Comparison of effective areas of a multilayer-coated Wolter-I mirror, as in Fig.~\ref{fig:comp_infty}, but with the source at a finite distance ($\delta$ = 5 arcmin, $D \approx $102 m). The error bars of the ray-tracing outputs are not shown. The accord between the curves is within a few percent on-axis (triangles) and 4~arcmin off-axis (circles).}
\label{fig:comp_finite}
\end{figure}

The results of this test are shown in Fig.~\ref{fig:comp_infty} for a source at infinity, and in Fig.~\ref{fig:comp_finite} for a source at a $D \simeq$ 102 m distance from the mirror. The results for a $10^4$ ray tracing for each energy value (1 keV steps) are plotted as symbols, whereas the results for the application of Eq.~(\ref{eq:Aeff_fin_offaxis0}) are plotted as lines. The statistical error of the ray-tracing results is of a few percent at low energies and close to 8\% at high energies, where the reflectivity is lower. We note the excellent matching of peak positions and shapes in both cases. A mismatch of a few percent can be observed at low energies, even at $\delta = \theta =0$, which is probably still related to the double cone approximation (Eq.~(\ref{eq:F_var})). 

From these examples, we also note that, even on-axis, the effective area is heavily reduced by the source at a 102~m distance from the mirror. This occurs because $\delta$ = 0.083 deg is close to $\alpha_0$~=~0.106 deg, although still smaller, thus the on-axis geometric area is only 12\% of the one we would have with $D = \infty$ (see Eq.~(\ref{eq:vign})). For the case $\delta$ = 0, the effective area also decreases with $\theta$ almost everywhere, as predicted by Eq.~(\ref{eq:Ageom_SC3}). The opposite effect is observed for a source at a 102~m distance, in agreement with Eq.~(\ref{eq:Ageom_D7}), which predicts an increase in the geometric area with $\theta$ when $\delta>\alpha_0/2$, as in the case that we considered. Finally, we also note how the reflectance features are {\it smoothed out} when the source moves off-axis, because of the variation in incidence angles over the reflecting surfaces.

\section{Final remarks and conclusions}
\label{Final}
We have shown how the problem of computing the effective area of a Wolter-I mirror with $f\ \gg L$ can be reduced to the computation of an integral (Eq.~(\ref{eq:Aeff_fin_offaxis}), or Eq.~(\ref{eq:Aeff_fin_offaxis0}) as a particular case), on the only condition that we are able to compute $r_{\lambda}(\alpha)$, the mirror reflectivity as a function of the photon wavelength and the incidence angle. This can easily be achieved numerically for a source at infinity, as in astronomical cases, or for a source at finite distance, as usually done for on-ground calibrations. For the ideal case of a constant reflectivity $r =1$, we could solve the integral and obtain {\it algebraic expressions for the geometric area} of the mirror. Finally, we presented some examples of the application of the formalism, and checked that its predictions agree with those of a detailed ray-tracing routine.

The analytical approach undoubtedly has several advantages. In general, the multilayer reflectivity computation is the Achilles' heel of a ray-tracing program that is aimed at determining the off-axis effective area of a mirror: although conceptually simple, the reflectivity computation consists of summing up the contribution of several layers to the reflectivity, so it requires a significant amount of computation time. A ray-tracing routine often requires $10^4 \div 10^5$ photons to return sufficient statistics for each photon energy, and the reflectivity routine has to be called for each of them. Hence, the total computation time can reach several hours. If the optic is still to be designed, the entire simulation needs to be run several times, adjusting the parameter values every time, until the optimal solution is reached: the optimization process can thereby take several days. 

In contrast, the analytical approach presented in this work is completely unaffected by statistical errors. Its accuracy is limited only by the applicability of the double cone approximation (discussed in Sect.~\ref{DC_WI}) and by the accuracy of the computation of the integral in Eq.~(\ref{eq:Aeff_fin_offaxis}). In practice, the reflectivity can be computed with, say, a 5 arcsec step of the incidence angle, without noticeably affecting the reflectivity of the multilayer. This means that, even for a very large off-axis angle, e.g., 10 arcmin, we need to compute the multilayer reflectivity only $\sim 500$ times, {\it at most}, instead of more than $10^4$ times as required by a ray-tracing.

For these reasons, the approach presented in this work might be extensively used to compute the effective area of grazing-incidence Wolter-I astronomical mirrors, on- and off-axis, whenever applicable. On the other hand, as already stated in Sect.~\ref{Intro}, it would {\it not} be applicable to very short focal lengths, or to systems of densely nested mirror shells, obstructing each other in the field of view: in this case a ray-tracing is, as of today, the only viable computational technique. Clearly, a ray-tracing is always necessary to investigate the angular resolution, especially whenever mirror deformations are present. 

We note that, if analytical expressions for $r_{\lambda}(\alpha)$ were available, it would be possible to solve explicitly the integral in Eq.~(\ref{eq:Aeff_fin_offaxis}) in the most general case, and obtain {\it algebraic formulae for the effective} area of a Wolter-I mirror. In this respect, an analytical approach for obtaining $r_{\lambda}(\alpha)$ of a multilayer was developed by Kozhevnikov~et~al.~(\cite{Kozhevnikov}), but a concise expression for the reflectivity appears still to be unavailable. However, if this is achievable, the computation of the effective area, and consequently the optical design and optimization, for {\it any} astronomical X-ray mirrors of sufficiently large $f\#$ might be simply reduced to the application of a handful of algebraic equations.

\appendix

\section{Vignetting for double reflection -- on-axis source at finite distance}
\label{App_vign}
We derive Eq.~(\ref{eq:vign}), which returns the geometrical vignetting of a double-reflection mirror, caused by the finite distance of the source. We consider a radial section of a Wolter-I mirror (Fig.~\ref{fig:mirror_section}), in the $xz$ plane, with the primary and secondary segment surfaces intersecting at a $2\alpha_0$ angle. We suppose that, as far as the effective area is concerned, the mirror profile can be approximated by a double cone. If the source were at infinity and on-axis (Fig.~\ref{fig:mirror_section}a), the incoming rays would impinge the primary and secondary segment at $\alpha_0$. We now move the X-ray source to a finite, although large, distance $D$. The beam is no longer collimated, since it has a nonzero half-divergency $\delta \approx R_0/D$, where $R_0$ is the radius at the intersection plane ($z=0$) of the mirror. Because of the small cross-section of the mirror as seen by the distant source, $\delta$ can be considered as a constant.

In this configuration, inspection of Fig.~\ref{fig:mirror_section}b shows that the incidence angle on the primary segment becomes $\alpha_0+\delta$, and that on the secondary one, $\alpha_0-\delta$. We consider a ray striking the primary segment at $A~=~(-R_0-z_0\alpha_0, 0, z_0)$, with $z_0 <L_1$. The angle formed by the ray direction with the optical axis, after the first reflection, is $2\alpha_0+\delta$: the equation of the reflected ray is then 
\begin{equation}
	z =- \frac{x+ R_0+z_0\alpha_0}{2\alpha_0+\delta}+z_0,
	\label{eq:exit_ray}
\end{equation}
where, as usual, we approximate the $\tan$ functions with the respective small arguments. If $z_0$ is sufficiently large, the ray misses the second reflection (Fig.~\ref{fig:mirror_section}b). The last reflection on the secondary mirror segment occurs if the reflected ray passes by the point $B= (-R_0+3\alpha_0 L_2, 0, -L_2)$. Substituting these coordinates into Eq.~(\ref{eq:exit_ray}), and solving for $z_0$, we obtain $Z_0$, the maximum value of $z_0$ for which we have a double reflection
\begin{equation}
	Z_0 \simeq \frac{\alpha_0-\delta}{\alpha_0+\delta}L_2,
	\label{eq:exit_ray3}
\end{equation}
regardless of $R_0$. Then the fraction of the primary segment that is {\it effective} for double reflection is $V = Z_0/L_1$, i.e.,
\begin{equation}
	V \simeq \frac{L_2(\alpha_0-\delta)}{L_1(\alpha_0+\delta)}.
	\label{eq:final_vign}
\end{equation}
We now consider the entire mirror, obtained by a rotation of the profile in Fig.~\ref{fig:mirror_section} about the optical axis. If the source is still on-axis, at a distance $D$, the vignetting given by Eq.~(\ref{eq:final_vign}) is easily applicable to all sectors of the mirror, and we obtain Eq.~(\ref{eq:vign}) exactly. In Appendix~\ref{App_angl}, we see that this result can easily be generalized to a source off-axis.

\section{Incidence angles and vignetting for double reflection -- detailed calculation for a source off-axis at finite distance}
\label{App_angl}
\begin{figure}
	\resizebox{\hsize}{!}{\includegraphics{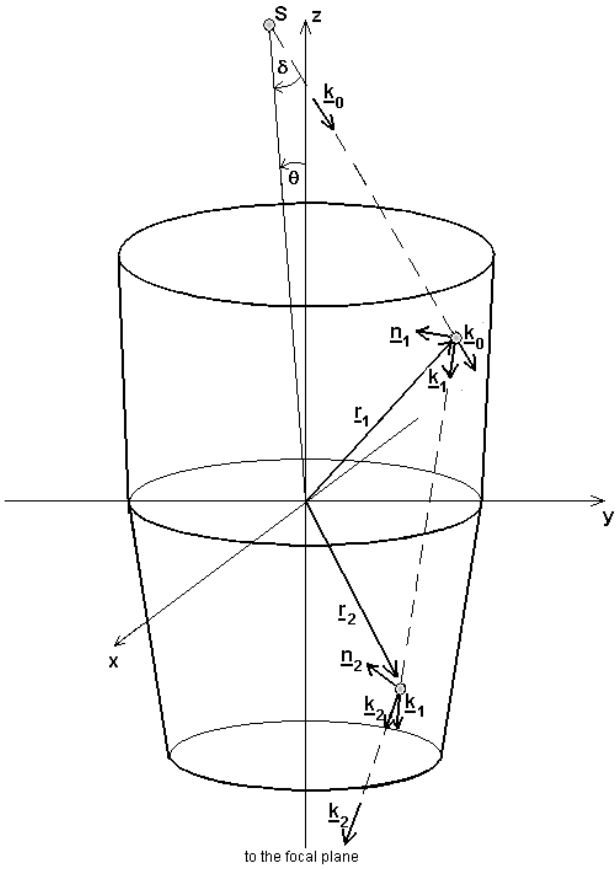}}
	\caption{Double reflection of an off-axis ray (dashed line) on a double cone mirror. Also shown are the direction vectors of the ray, and the normal vectors to the surface, $\underline{n}_1$ and $\underline{n}_2$, at the impact points, $\underline{r}_1$ and $\underline{r}_2$.}
\label{fig:offax_scheme}
\end{figure}

We consider a double cone mirror (Fig.~\ref{fig:offax_scheme}), with optical axis aligned with $z$ and the intersection plane at $z=0$. We define $R_0$ to be its radius at $z =0$, $\alpha_0$ the incidence angle for an on-axis source placed at infinity, and $L_1$, $L_2$, lengths of the primary and secondary mirror segments. We assume $\alpha_0$ to be shallow and that the source is at the finite distance $D \gg L_1$. If the source is on-axis, the beam impinges the primary segment with a nearly constant half-divergency $\delta = R_0/D$. We denote with $r_1$, $\varphi_1$, and $z_1$ the radial, azimuthal, and axial coordinates of the impact point on the primary segment, and $r_2$, $\varphi_2$, and $z_2$ on the secondary. We now assume the source to be moved off-axis by an angle $\theta$, and choose the direction $\varphi_1 = \varphi_2 = 0$ in the tilt plane of the source (the $xz$ plane, like in Fig.~\ref{fig:mirror_section}). Our scope in this appendix is to determine analytically the incidence angles on the two mirrors, $\alpha_1$ and $\alpha_2$, and the vignetting for double reflection, $V$, as a function of $\alpha_0$, $\delta$, $\theta$, $\varphi_1$, more generally than we did in Appendix~\ref{App_vign}. 

The two conical surfaces are described in polar coordinates by the equations
\begin{eqnarray}
	z_1 = \frac{r_1-R_0}{\alpha_0} & \mbox{with $r_1> R_0$}, \label{eq:z1}\\
	z_2 = \frac{r_2-R_0}{3\alpha_0} & \mbox{with $r_2 < R_0$}.\label{eq:z2}
\end{eqnarray}
The normal vectors to the two segments, directed inwards, are 
\begin{eqnarray}
	\underline{n}_1 = \left(
	\begin{array}{c}
	-\cos\alpha_0\cos\varphi_1\\
	-\cos\alpha_0\sin\varphi_1\\
	\sin\alpha_0
	\end{array}
	\right), &
	\underline{n}_2 = \left(
	\begin{array}{c}
	-\cos3\alpha_0\cos\varphi_2\\
	-\cos3\alpha_0\sin\varphi_2\\
	\sin3\alpha_0
	\end{array}
	\right);
\label{eq:normal}
\end{eqnarray}
if the double cone approximation is valid, the normal vectors are only a function of the $\varphi$'s angles. {\it If} the source were on-axis, the initial direction of the ray would have the expression
\begin{equation}
	\underline{k}_0^* = \left(
	\begin{array}{c}
	\sin\delta\cos\varphi_1\\
	\sin\delta\sin\varphi_1\\
	-\cos\delta
	\end{array}
	\right).
\label{eq:k_0star}
\end{equation}
Since the source is off-axis by $\theta$, we have to tilt $\underline{k}_0^*$ by an angle $\theta$ about the $y$ axis. The application of the rotation matrix returns the expression for the initial direction of the off-axis photon, $\underline{k}_0$,
\begin{equation}
	\underline{k}_0 = \left(
	\begin{array}{c}
	\sin\delta\cos\varphi_1\cos\theta-\cos\delta\sin\theta\\
	\sin\delta\sin\varphi_1\\
	-\sin\delta\cos\varphi_1\sin\theta-\cos\delta\cos\theta
	\end{array}\right).
\label{eq:k_0}
\end{equation}
If $\alpha_1$ is the incidence angle of the first reflection (measured from the surface), we can write the scalar product 
\begin{equation}
	\cos\left(\frac{\pi}{2}+\alpha_1\right) = \underline{k}_0\cdot\underline{n}_1,
\label{eq:scal_prod}
\end{equation}
which becomes, after some algebra,
\begin{eqnarray}
	\sin\alpha_1 &=& \cos\alpha_0\sin\delta\,(\cos\theta\cos^2\varphi_1+\sin^2\varphi_1) +\nonumber\\
	&-&\cos(\alpha_0+\delta)\sin\theta\cos\varphi_1 + \cos\delta\cos\theta\sin\alpha_0.
\label{eq:theta_p}
\end{eqnarray}
In the limit of small $\alpha_0$, $\delta$, $\theta$, we can approximate the cosines with 1 and the sines with their arguments, yielding
\begin{equation}
	\alpha_1 \simeq \alpha_0+\delta-\theta\cos\varphi_1, 
\label{eq:theta_p_appr}
\end{equation}
that is exactly Eq.~(\ref{eq:angles1}).

To derive the incidence angle on the secondary mirror segment, we need to trace the exit direction of each ray, $\underline{k}_1$, after the first reflection. This is obtained from the vector equation 
\begin{equation}
	 \underline{k}_1 = \underline{k}_0 - 2(\underline{k}_0\cdot\underline{n}_1)\, \underline{n}_1,
\label{eq:exit_dir}
\end{equation}
because the parallel component to the surface is conserved, whilst the normal component reverses its sign in the reflection process. By substitution of the Eqs.~(\ref{eq:k_0})~and~(\ref{eq:normal}), one obtains, as always in the small angles limit,
\begin{equation}
	 \underline{k}_1 \simeq \left(
	\begin{array}{c}
	-\theta+(\delta-2\alpha_1)\cos\varphi_1\\
	(\delta-2\alpha_1)\sin\varphi_1\\
	-1
	\end{array}\right).
\label{eq:exit_dir1}
\end{equation}
A ray reflected at the generic point $\underline{r}_1$ of the primary segment has equation $\underline{r}(t) = \underline{r}_1+ t \,\underline{k}_1$, with $t>0$. Therefore, the reflected ray intersects the secondary segment at a position $\underline{r}_2$ fulfilling the condition $ \underline{r}(t)=\underline{r}_2 $, i.e., $ \underline{r}_2 - \underline{r}_1 = t\, \underline{k}_1$ for some $t$. This is equivalent to constraining the two vectors to be parallel:
\begin{equation}
 	(\underline{r}_2 - \underline{r}_1)\times\underline{k}_1 = \underline{0},
\label{eq:prod_vect}
\end{equation}
where $\times$ denotes a cross product. 

Using both Eqs.~(\ref{eq:z1}) and~(\ref{eq:z2}), Eq.~(\ref{eq:prod_vect}) can be developed into 3 scalar equations, only 2 of which are mutually independent, i.e.,
\begin{eqnarray}
 	r_2\left(1+\frac{3\alpha_0}{k_{1x}}\cos\varphi_2\right) +2R_0 &=&3r_1\left(1+\frac{\alpha_0}{k_{1x}}\cos\varphi_1\right), \label{eq:equation1}\\
	r_2\left(1+\frac{3\alpha_0}{k_{1y}}\sin\varphi_2\right) +2R_0 &=& 3r_1\left(1+\frac{\alpha_0}{k_{1y}}\sin\varphi_1\right). \label{eq:equation2}
\end{eqnarray}
When $\theta = \delta =0$, Eq.~(\ref{eq:exit_dir1}) infers $k_{1x}~=~-2\alpha_0\cos\varphi_1$ and $k_{1y}~=~-2\alpha_0\sin\varphi_1$. In this case, the solution is barely $\varphi_2~=~\varphi_1$ and $r_2 = 4R_0-3r_1$. 

We now search for a perturbative solution of these equations, of the kind $\varphi_2 = \varphi_1+\varepsilon$ and $r_2 = 4R_0-3r_1+\xi$, with $\varepsilon~\ll~\varphi_1$ and $\xi~\ll~r_1$. Substituting these expressions into Eqs.~(\ref{eq:equation1}) and~(\ref{eq:equation2}), neglecting terms in $\xi\varepsilon$, and using Eq.~(\ref{eq:exit_dir1}) for the components of $\underline{k}_1$, we obtain a linear system in $\xi$ and $\varepsilon$, whose solution is
\begin{eqnarray}
 	\xi& \simeq &\frac{6(\alpha_0-\alpha_1)(r_1-R_0)}{2\alpha_0-\alpha_1} \label{eq:xi}\\
	\varepsilon & \simeq & \frac{2\theta\,\sin\varphi_1\,(r_1-R_0)}{(4R_0-3r_1)(2\alpha_0-\alpha_1)}.\label{eq:eps}
\label{eq:new_pos}
\end{eqnarray}
For small off-axis angles, $r_1 -R_0 \lesssim L\alpha_0$, $2\alpha_0-\alpha_1 \approx \alpha_0$, and $4R_0 -3r_1 \approx R_0$, $\varepsilon$ must then be of the order of $2\theta L/R_0$ or less, as expected from the simple argument presented in Sect.~\ref{offaxis}.

We can now derive the incidence angle on the secondary mirror segment, $\alpha_2$. To this end, we write the scalar product
\begin{equation}
	\cos\left(\frac{\pi}{2}+\alpha_2\right) = \underline{k}_1\cdot\underline{n}_2,
\label{eq:scal_prod2}
\end{equation}
where $\underline{n}_2$ is provided by Eq.~(\ref{eq:normal}). This would require the computation of $\cos(\varphi_1+\varepsilon)$ and $\sin(\varphi_1+\varepsilon)$, which would yield a complicated expression. Nevertheless, all terms in $\varepsilon$ are of second order, so they can be neglected in small angles approximation. We are then allowed to assume that $\varphi_1 \approx \varphi_2$ and obtain after some passages $\alpha_2 \simeq 2\alpha_0-\alpha_1$, that is,
\begin{equation}
	\alpha_2 \simeq \alpha_0-\delta+\theta\cos\varphi_1, 
\label{eq:theta_h_appr}
\end{equation}
so we have found Eq.~(\ref{eq:angles2}). This concludes the calculation of the incidence angles for an off-axis source. 

Finally, we proceed to compute the vignetting factor for double reflection, in the general case of an off-axis source. In the following, since $\varphi_1 \simeq \varphi_2$, we suppress the subscript and denote with $\varphi$ their nearly-common value. From the value of $\xi$ (Eq.~(\ref{eq:xi})), and using Eq.~(\ref{eq:theta_h_appr}), we derive $r_2$, 
\begin{equation}
	r_2 \simeq \frac{2(\alpha_0+\alpha_1)R_0-3\alpha_1r_1}{\alpha_2}.
	\label{eq:r2}
\end{equation}
Reflection on the secondary mirror segment occurs if $z_2 \geq -L_2$, i.e., $r_2 \geq R_0 -3\alpha_0 L_2$ (see Eq.~(\ref{eq:z2})). Substituting Eq.~(\ref{eq:r2}) into this inequality, using Eq.~(\ref{eq:z1}), and solving for $z_1$, we derive
\begin{equation}
	z_1(\varphi) \leq Z_0 = L_2\frac{\alpha_2(\varphi)}{\alpha_1(\varphi)},
	\label{eq:z2_final}
\end{equation}
where we indicate explicitly the dependence of the incidence angles on the azimuthal angle. So the fraction of the primary mirror segment effective for double reflection, at the polar angle~$\varphi$, is $Z_0/L_1$ or 
\begin{equation}
	V(\varphi) = \frac{L_2\,\alpha_2(\varphi)}{L_1\,\alpha_1(\varphi)}.
	\label{eq:vignetting_general}
\end{equation}
The last equation generalizes Eq.~(\ref{eq:final_vign}) to the case of a source off-axis: it is Eq.~(\ref{eq:vign_gen}), which we used for the computation of the off-axis effective area in Sect.~\ref{offaxis}.

\begin{acknowledgements}
This research is funded by ASI (the Italian Space Agency). The authors acknowledge for useful discussions G.~Pareschi, G.~Tagliaferri, S.~Campana, A.~Moretti (INAF/OAB).
\end{acknowledgements}

\end{document}